 \documentclass[pra,a4paper,oneside,twocolumn,superscriptaddress,showpacs]{revtex4-1}

\usepackage[english]{babel}
\usepackage[latin1]{inputenc}
\usepackage[T1]{fontenc}
\usepackage{lmodern}
\usepackage{ulem}
\usepackage{dcolumn}
\usepackage{subfigure}
\usepackage{multirow}
\usepackage{amsmath,amsfonts,amssymb}

\usepackage{graphicx}  
\usepackage{latexsym}   
\usepackage{color}

\begin{document}

\title{Tunable multiple Fano resonances in magnetic single-layered core-shell particles}

\author{\firstname{Tiago}  J. \surname{Arruda}}
\email{tiagoarruda@pg.ffclrp.usp.br}
\affiliation{Faculdade de Filosofia,~Ci\^encias e Letras de Ribeir\~ao
Preto, Universidade de S\~ao Paulo, 14040-901 Ribeir\~ao
Preto, S\~ao Paulo, Brazil}

\author{\firstname{Alexandre} S. \surname{Martinez}}
\affiliation{Faculdade de Filosofia,~Ci\^encias e Letras de Ribeir\~ao
Preto, Universidade de S\~ao Paulo, 14040-901 Ribeir\~ao
Preto, S\~ao Paulo, Brazil}
\affiliation{National Institute of Science and Technology in
Complex Systems, 22290-180 Rio de Janeiro, Rio de Janeiro, Brazil}

\author{\firstname{Felipe}  A. \surname{Pinheiro}}
\affiliation{Instituto de F\'{i}sica, Universidade Federal do Rio de Janeiro, 21941-972 Rio de Janeiro, Rio de Janeiro, Brazil}
\affiliation{Optoelectronics Research Centre and Centre for Photonic Metamaterials, University of Southampton, 
Highfield, Southampton SO17 1BJ, United Kingdom}

\begin{abstract}

We investigate multiple Fano, comblike scattering resonances in single-layered, concentric core-shell nanoparticles composed of magnetic materials. 
Using the Lorenz-Mie theory, we derive, in the long-wavelength limit, an analytical condition for the occurrence of comblike resonances  in the single scattering by coated spheres. 
This condition establishes that comblike scattering response uniquely depends on material parameters and thickness of the shell, provided that it is magnetic and thin compared to the scatterer radius.
We also demonstrate that comblike scattering response shows up beyond the long-wavelength limit and it is robust against absorption.
Since multiple Fano resonances are shown to depend explicitly on the magnetic permeability of the shell, we argue that both the position and profile of the comblike, morphology-dependent resonances could be externally tuned by exploiting the properties of engineered magnetic materials.  

\end{abstract}

\pacs{
		 42.25.Fx,  
     42.79.Wc, 
     41.20.Jb,   
     78.20.Ci    
}

\maketitle


\section{Introduction}

Optical systems exhibiting comblike resonances have a wide range of scientific and technological applications, especially in spectroscopy and frequency metrology~\cite{haye2007,haye2011,schliesser}.
Comblike profiles, which consist of ultrasharp resonance peaks, can be achieved {\it e.g.} in the electromagnetic (EM) scattering by plasmonic nanoparticles or via morphology-dependent resonances in nonmetallic scatterers.
Indeed, plasmonic nanoparticles support localized surface plasmon resonances that are strongly dependent on their geometrical and material parameters, allowing for engineering the scattering response~\cite{zayats-plasmon}. 
In addition, scatterers with a high degree of symmetry, such as spheres, spheroids and cylinders, may support morphology-dependent resonances, which are related to constructive interferences confined inside the particle by almost total internal reflections~\cite{johnson}.

Recently, a great deal of attention has been devoted to Fano resonances in plasmonic nanostrutures due to its sensitivity to both geometry and local environment changes~\cite{luk,kivshar,halas2010}.
The interference between a broad bright resonance and a narrow dark resonance modes, supported by plasmonic nanostrutures, gives rise to the Fano resonance, which has a characteristic, asymmetric lineshape with a narrow bandwidth~\cite{kivshar}. 
By properly choosing the materials and/or the system design, it is possible to generate {\it e.g.} multiple Fano, comblike scattering resonances in plasmonic nanoparticles~\cite{artar2011,fu2012,monticone2013,zayats}.
The design of these Fano-comb nanoparticles may enable several applications at different spectral ranges, such as improving the resolution in comb-spectroscopy techniques~\cite{schliesser}, and optical tagging~\cite{monticone2013}. 

One possible way to achieve comblike resonances in EM scattering is to consider multilayered core-shell nanoparticles~\cite{monticone2013}. 
The main idea is to combine localized plasmon resonances and the scattering cancellation technique for a single core-shell scatterer~\cite{alu2005,alu2008,argyro,wilton1,wilton2,wilton3}, whose core is dielectric and the coating is composed of multilayered semiconductor materials with a gradient of doping level.  
An alternative approach to obtain Fano-comb resonances is to consider nonconcentric, single-layered cylindrical  nanostructures~\cite{zayats}, whose fabrication is considerably simpler than multilayered nanoparticles. 

In the present paper, we propose an alternative approach, based on magnetic materials, for generating comblike scattering response in concentric, single-layered core-shell nanoparticles, which can exhibit Fano resonances~\cite{tiago-pra,chen,jordi,rybin}. 
Instead of inducing Fano-comb resonances by the geometrical symmetry breaking in single-layered core-shell nanostructures~\cite{zayats}, we suggest the use of high permeability covers~\cite{prl1} in center-symmetric scatterers. 
Indeed, in the long-wavelength limit, the presence of a ferromagnetic layer breaks the isotropy of the Rayleigh scattering, and allow us to obtain multiple Fano, comblike scattering resonances composed of both electric and magnetic dipole resonances in the extinction cross-sections~\cite{tiago-sphere-cylinder}.
These Fano-like resonances, known as ``unconventional'' Fano resonances~\cite{tribelsky}, are  different from the conventional Fano ones~\cite{luk}, since they are polarization-independent and arise from the interference between EM modes with the same multipole moment.

We explore these properties to derive analytical conditions for the occurrence of comblike resonances in coated spheres composed of a dielectric core and a high magnetic permeability shell.
As the condition for the occurrence of multiple Fano resonances depends explicitly on the value of the magnetic permeability of the shell, we argue that both the position and profile of the comblike resonances could be externally tuned.
This tuning could be achieved by exploiting {\it e.g.} the properties of engineered magnetic materials, where the magnetic permeability can be modified by applying an external magnetic field and/or by varying the temperature. 

This paper is organized as follows.
In Sec.~\ref{theory}, we briefly review the Lorenz-Mie theory for center-symmetric coated spheres~\cite{bohren,tiago-joa}. 
By applying the long-wavelength approximation, we obtain in Sec.~\ref{analytical} an analytical condition to determine the position of the multiple Fano, comblike scattering resonances as a function of the shell parameters.  
In Sec.~\ref{numerical}, we go beyond the long-wavelength limit and consider a system composed of a dielectric nanosphere coated with a magnetic thin shell whose permittivity is provided by a realistic, lossy Drude model.
Finally, in Sec.~\ref{conclusion}, we summarize our main results and conclude. 

\section{Theoretical background: Lorenz-Mie formalism}
\label{theory}

\begin{figure}[htpb]
\centerline{\includegraphics[width=.7\columnwidth]{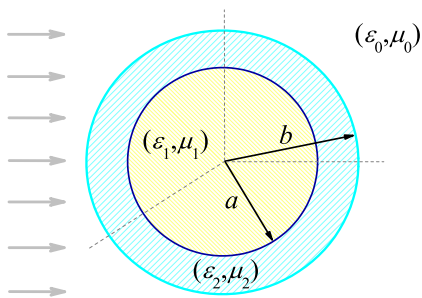}}
\caption{(Color online) A core-shell scatterer with spherical symmetry irradiated by a plane wave.
The core has optical properties $(\varepsilon_{1},\mu_{1})$ and radius $a$, whereas the shell has $(\varepsilon_{2},\mu_{2})$ and radius $b$.
The surrounding medium is the vacuum $(\varepsilon_0,\mu_0)$. }\label{fig1}
\end{figure}

Here we present the main results of the analytic solution of EM scattering by coated spheres, kwown as the Lorenz-Mie theory~\cite{bohren}, that are  extensively used throughout this paper.  
Let us assume for the incident EM wave the time harmonic dependence $e^{-\imath\omega t}$, with $\omega$ being the angular frequency.
The scatterer is a center-symmetric, core-shell sphere, with inner radius $a$ and outer radius $b$, composed of linear, spatially homogeneous, and isotropic magneto-dielectric materials.
The electric permittivities $(\varepsilon)$ and magnetic permeabilities ($\mu$) of core ($0< r\leq a$), shell ($a< r\leq b$) and embedding medium $(b < r)$ are $(\varepsilon_{1},\mu_{1})$, $(\varepsilon_{2},\mu_{2})$ and $(\varepsilon_0,\mu_0)$, respectively.
The scatterer geometry is depicted in Fig.~\ref{fig1}.

For the sake of simplicity, the medium $(\varepsilon_0,\mu_0)$ is considered to be the vacuum and the relative parameters are $\varepsilon_{q{\rm r}}\equiv\varepsilon_{q}/\varepsilon_0$ and $\mu_{q{\rm r}}\equiv\mu_{q}/\mu_0$, with $q=1$ for the core and $q=2$ for the shell.
The scattering efficiency (which is the respective cross-section in units of $\pi b^2$) is
\begin{align}
Q_{\rm sca}=\frac{2}{y^2}\sum_{n=1}^{\infty}(2n+1)\left(|a_n|^2+|b_n|^2\right)\ ,\label{Qsca}
\end{align}
where $y=kb$ is the size parameter of the outer sphere ($k=2\pi/\lambda$ being the incident wave number)~\cite{bohren}.
The extinction and absorption efficiencies are, respectively, $Q_{\rm ext}=({2}/{y^2})\sum_{n=1}^{\infty}(2n+1){\rm Re}(a_n+b_n)$ and $Q_{\rm abs}=Q_{\rm ext}-Q_{\rm sca}$.
Here, the Lorenz-Mie scattering coefficients $a_n$ (electric) and $b_n$ (magnetic) are~\cite{bohren,tiago-joa}:
\begin{align}
    a_n&=\frac{\left(\widetilde{D}_n/\widetilde{m}_2+n/y\right)\psi_n(y)-\psi_{n-1}(y)}{\left(\widetilde{D}_n/\widetilde{m}_2+n/y\right)\xi_n(y)-\xi_{n-1}(y)}\
    ,\label{an}\\
    b_n&=\frac{\left(\widetilde{m}_2\widetilde{G}_n+n/y\right)\psi_n(y)-\psi_{n-1}(y)}{\left(\widetilde{m}_2\widetilde{G}_n+n/y\right)\xi_n(y)-\xi_{n-1}(y)}\ ,
\end{align}
where one defines the auxiliary functions~\cite{bohren}
\begin{align}
    \widetilde{D}_n&=\frac{D_n(m_2y)-A_n\chi_n'(m_2y)/\psi_n(m_2y)}{1-A_n\chi_n(m_2y)/\psi_n(m_2y)}\
    ,\label{Dn}\\
    \widetilde{G}_n&=\frac{D_n(m_2y)-B_n\chi_n'(m_2y)/\psi_n(m_2y)}{1-B_n\chi_n(m_2y)/\psi_n(m_2y)}\
    ,\\
    A_n&=\frac{\psi_n(m_2x)\left[\widetilde{m}_2D_n(m_1x)-\widetilde{m}_1D_n(m_2x)\right]}{\widetilde{m}_2D_n(m_1x)\chi_n(m_2x)-\widetilde{m}_1\chi_n'(m_2x)}\
    ,\label{An}\\
    B_n&=\frac{\psi_n(m_2x)\left[\widetilde{m}_2D_n(m_2x)-\widetilde{m}_1D_n(m_1x)\right]}{\widetilde{m}_2\chi_n'(m_2x)-\widetilde{m}_1D_n(m_1x)\chi_n(m_2x)}
    ,\label{Bn}
\end{align}
with $x=ka$ being the size parameter of the inner sphere and $D_n(\rho) \equiv {\rm d}[\ln \psi_n(\rho)]/{\rm d}\rho$.
The functions $\psi_n(\rho)=\rho j_n(\rho)$, $\chi_n(\rho)=-\rho y_n(\rho)$ and $\xi_n(\rho)=\psi_n(\rho)-\imath\chi_n(\rho)$ are the Riccati-Bessel, Riccati-Neumann and Riccati-Hankel functions, respectively, with $j_n$ and $y_n$ being the spherical Bessel and Neumann functions~\cite{bohren}.
The relative refractive index is $m_q=\sqrt{\varepsilon_{q{\rm r}}\mu_{q{\rm r}}}$ and $\widetilde{m}_q \equiv \sqrt{\varepsilon_{q{\rm r}}/\mu_{q{\rm r}}}$, with $q=\{1,2\}$~\cite{prl1}.
Notice that $\widetilde{m}_q$ is the inverse of the usual definition of relative impedance, $Z_{q{\rm r}}=\sqrt{\mu_{q{\rm r}}/\varepsilon_{q{\rm r}}}$, so that $\widetilde{m}_q=m_q$ for nonmagnetic materials $(\mu_{q{\rm r}}=1)$~\cite{prl1}.

Equations~(\ref{Qsca})--(\ref{Bn}) are the exact expressions for core-shell spherical scatterers with arbitrary geometrical and material parameters.
In the following, we consider approximations for both core and shell parameters regarding the incident wavelength.

\section{Analytical results}
\label{analytical}

We proceed our analysis by taking into account the long-wavelength regime $(y=kb\ll1)$.
Within this approximation, the arguments $m_1x$ (core) and $m_2y$ (shell) of spherical Bessel and Neumann functions should not necessarily be small.
Following Tribelsky {\it et al.}~\cite{tribelsky}, one can write $a_n ={\mathcal{F}_n^{(A)}}/[{\mathcal{F}_n^{(A)}+\imath\mathcal{G}_n^{(A)}}]$ and $b_n={\mathcal{F}_n^{(B)}}/[{\mathcal{F}_n^{(B)}+\imath\mathcal{G}_n^{(B)}}]$.
For coated spheres, we calculate the new auxiliary functions in the small-particle limit; they read
\begin{align}
\mathcal{F}_n^{(A)}&\approx\frac{y^n}{(2n+1)!!}\left[(n+1)\widetilde{m}_2-y\widetilde{D}_n\right]\ ,\label{Fn-te}\\
\mathcal{G}_n^{(A)}&\approx\frac{(2n-1)!!}{y^n}\left[\widetilde{D}_n+\frac{\widetilde{m}_2n}{y}\right]\ ,\label{Gn-te}\\
\mathcal{F}_n^{(B)}&\approx\frac{y^n}{(2n+1)!!}\left[\widetilde{m}_2y\widetilde{G}_n-(n+1)\right]\nonumber\\
&\ -\frac{y^{n+2}}{2(2n+3)!!}\left[\widetilde{m}_2y\widetilde{G}_n-(n+3)\right]\ ,\label{Fn-tm}\\
\mathcal{G}_n^{(B)}&\approx-\frac{{(2n-1)!!}}{y^n}\left[\widetilde{m}_2\widetilde{G}_n+\frac{n}{y}\right]\ .\label{Gn-tm}
\end{align}

We find that there are electric and magnetic scattering resonances ({\it i.e.}, constructive interferences) in $|a_n|^2$ and $|b_n|^2$ when $\mathcal{G}_n^{(A,B)}=0$ ($|a_n|^2=|b_n|^2=1$), respectively.
Similarly, destructive interferences occur whenever $\mathcal{F}_n^{(A,B)}=0$ ($|a_n|^2=|b_n|^2=0$). 
These destructive interferences lead to very small scattering cross-sections, and characterize an EM ``cloaking'' mode. 
Note that the relevant variables to analyze electric and magnetic resonances are indeed $\mathcal{G}_n^{(A)}/\mathcal{F}_n^{(A)}$ and $\mathcal{G}_n^{(B)}/\mathcal{F}_n^{(B)}$, respectively~\cite{johnson}.
For nonmagnetic particles $(\mu_{1{\rm r}}=\mu_{2{\rm r}}=1)$ the leading term in Eq.~(\ref{Fn-tm}) vanishes, so that it requires higher accuracy relative to other approximations~\cite{tribelsky}.
In particular, the expressions presented in Ref.~\cite{tribelsky} for a homogeneous sphere are retrieved as one performs the following mapping: $a\to b$ [$\widetilde{D}_n=D_n(m_1x)\widetilde{m}_2/\widetilde{m}_1$, $\widetilde{G}_n=D_n(m_1x)\widetilde{m}_1/\widetilde{m}_2$] or $a\to0$ or $m_1=m_2$ [$\widetilde{D}_n=\widetilde{G}_n=D_n(m_2y)$].

\subsection{Shell in the ferromagnetic limit ($|m_2|y\gg1$, $y\ll1$)}

In addition to the small-particle limit ($y\ll1$), if $|m_2|y\gg 1$ ({\it i.e.}, large permittivity and/or large permeability shell materials), one has $\psi_n(m_2y)\approx\sin(m_2y-n\pi/2)$ and $\chi_n(m_2y)\approx\cos(m_2y-n\pi/2)$~\cite{bohren}.
To simplify our analysis, let us define the ``phase shifts'' 
\begin{align}
	\Delta_n^{(A)}&=\tan^{-1}(-A_n)\ ,\\
	\Delta_n^{(B)}&=\tan^{-1}(-B_n)\ ,
\end{align}
where $A_n$ and $B_n$ are defined in Eqs.~(\ref{An}) and (\ref{Bn}), respectively.
The ferromagnetic limit ($|m_2|y\gg1$) is obtained from Eqs.~(\ref{Fn-te})--(\ref{Gn-tm}) by replacing $\widetilde{D}_n$ with $\cot[m_2y+\Delta_n^{(A)}-n\pi/2]$ and $\widetilde{G}_n$ with $\cot[m_2y+\Delta_n^{(B)}-n\pi/2]$, and redefining the auxiliary functions.
In particular, the dipole approximation ($n=1$) for the ferromagnetic small-particle limit leads to:
\begin{align}
\mathcal{F}_1^{(A)}&\approx\frac{y}{3}\left\{2\widetilde{m}_2+y\tan\left[m_2y+\Delta_1^{(A)}\right]\right\}\ ,\label{F1-te}\\
\mathcal{G}_1^{(A)}&\approx\frac{1}{y}\left\{\frac{\widetilde{m}_2}{y}-\tan\left[m_2y+\Delta_1^{(A)}\right]\right\}\ ,\label{G1-te}\\
\mathcal{F}_1^{(B)}&\approx-\frac{y}{3}\left\{\widetilde{m}_2y+2\cot\left[m_2y+\Delta_1^{(B)}\right]\right\}\ ,\label{F1-tm}\\
\mathcal{G}_1^{(B)}&\approx\frac{{1}}{y}\left\{\widetilde{m}_2-\frac{1}{y}\cot\left[m_2y+\Delta_1^{(B)}\right]\right\}\ .\label{G1-tm}
\end{align}
In Eqs.~(\ref{F1-te})--(\ref{G1-tm}), the dependence on the core parameters $(\varepsilon_{1},\mu_{1})$ and $x=ka$ is encoded in $\Delta_1^{(A,B)}$.
Regarding the shell parameters $(\varepsilon_{2},\mu_{2})$, we consider two reciprocal configurations that fulfill the limiting case of a shell with high refractive index ($|m_2|=|\sqrt{\varepsilon_{2{\rm r}}\mu_{2{\rm r}}}|\gg1$), namely:
$(i)$ large permeability shells with small permittivity ($|\varepsilon_{2{\rm r}}|\ll1;|\mu_{2{\rm r}}|\gg1$) and
$(ii)$ large permittivity shells with small permeability ($|\varepsilon_{2{\rm r}}|\gg1;|\mu_{2{\rm r}}|\ll1$).
By symmetry, one case can readily be obtained from the other by replacing $\varepsilon$ with $\mu$; for our purposes, we focus on case $(i)$, {\it i.e.}, high permeability materials with small permittivities such that $|\mu_{2{\rm r}}|\gg|\varepsilon_{2{\rm r}}|$.

We recall that vanishing shell parameters facilitate the scattering cancellation (EM cloaking).
This mechanism is based on the cancellation of the EM fields due to a local negative polarizability coefficient.
The electric dipole ($n=1$) is associated with the local polarization vector $\mathbf{P}(\mathbf{r},\omega)=\varepsilon_0[\varepsilon_{\rm r}(\mathbf{r},\omega)-1]\mathbf{E}(\mathbf{r},\omega)$, where one considers the local fields and parameters inside the scatterer.
Thereby, in case $(i)$, one may locally induce a polarization vector out of phase to the electric field, allowing for the partial cancellation of the scattering efficiency ($Q_{\rm sca}\approx0$).
In addition, large values of the magnetic permeability in small-particle limit are responsible for breaking the isotropy of the Rayleigh scattering, favoring backscattering~\cite{prl1} and EM resonances inside the particle~\cite{tiago-sphere-cylinder}.
Consequently, one could in principle engineer suitable properties to tune both cloaking and strong scattering responses in magnetic core-shell spheres.

In case~$(i)$, spherical shell materials with $|\mu_{2{\rm r}}|\gg1$ lead to high refractive ($|m_2|\gg1$) and low impedance ($|\widetilde{m}_2|\ll1$) indices, since low permittivity is considered.
Therefore, from Eqs.~(\ref{F1-te}) and (\ref{F1-tm}), the electric and magnetic cloaking in the ferromagnetic limit occur for $\tan[m_2y+\Delta_1^{(A)}]\approx 0$ and $\cot[m_2y+\Delta_1^{(B)}]\approx 0$, respectively.
Conversely, for case $(ii)$, $|\varepsilon_{2{\rm r}}|\gg1$ leads to $|m_2|\gg1$ and $|\widetilde{m}_2|\gg1$, which readily provide the electric and magnetic cloaking conditions: $\cot[m_2y+\Delta_1^{(A)}]\approx0$ and $\tan[m_2y+\Delta_1^{(B)}]\approx0$.
These are the analytical conditions for EM cloaking to occur in both cases $(i)$ and $(ii)$.
Notice that, for this set of cloaking conditions, one can readily obtain case $(i)$ from case $(ii)$ by replacing $A_n$ with $B_n$, and vice versa.
In the following, we present a special configuration where multiple electric and magnetic Fano resonances in the scattering cross-section are achieved.

\subsection{Core in the Rayleigh limit ($|m_1|x\ll1$, $x\ll1$)}

Let us consider now a non-dispersive dielectric core $(\varepsilon_{1},\mu_{1})$ coated with a dispersive magnetic shell $[\varepsilon_{2}(\omega),\mu_{2}(\omega)]$, with $|\mu_{2{\rm r}}(\omega)|\gg1$, both of them lossless media.
Fixing the size parameters $x=ka$ and $y=kb$, the cloaking condition at resonance $\omega=\omega_0$ [$Q_{\rm sca}(\omega_0)=0$] takes place for a certain refractive index $\overline{m}_2\equiv m_2(\omega_{0})$ and $\overline{\Delta}_1^{(A,B)}\equiv{\Delta}_1^{(A,B)}(\omega_0)$: 
\begin{align}
\tan\left[\overline{m}_2y+\overline{\Delta}_1^{(A)}\right]=\cot\left[\overline{m}_2y+\overline{\Delta}_1^{(B)}\right]= 0\ .
\end{align}
Since the arguments of these trigonometric functions are very large, it is convenient to consider a small frequency variation $\delta\omega$ in the vicinity of $\omega_0$, leading to a corresponding variation $\delta m$ (which is not necessarily small) in the refractive index: $m_2(\omega_{0}+\delta\omega)=\overline{m}_2+\delta m$.
To simplify our analysis, consider that only $\varepsilon_{2}=\varepsilon_{2}(\omega)$ depends on the frequency in this range $\delta\omega$, so that the impedance index remains approximately unchanged: $\widetilde{m}_2(\omega_0+\delta\omega)\approx\widetilde{m}_2(\omega_0)=\widetilde{m}_2\ll1$.

In addition to the small-particle $(x<y\ll1)$ and ferromagnetic $(|m_2|y\gg1)$ limits for the shell, we impose that the dielectric core satisfies the Rayleigh limit: $|m_1|x\ll1$.
This provides $A_1\approx [\varepsilon_{1{\rm r}}x\tan(m_2x)+2\widetilde{m}_2]/[\varepsilon_{1{\rm r}}x-2\widetilde{m}_2\tan(m_2x)]$, where we have used $D_1(m_1x)\approx2/(m_1x)+\mathcal{O}(x)$ and $B_1$ is readily obtained from $A_1$ by replacing ($\varepsilon_{1{\rm r}},\widetilde{m}_2$) with $(\mu_{1{\rm r}},1/\widetilde{m}_2)$.
Consequently, $\Delta_1^{(A,B)}(\omega_0+\delta\omega)\approx\overline{\Delta}_1^{(A,B)}-(\delta m) x$.
From the latter approximation, we can finally write the analytical expressions in terms of the aspect ratio $S\equiv a/b=x/y$, which is a geometric parameter that plays a crucial role on the EM cloaking.
Imposing $S\approx1$, we obtain 
\begin{align*}
\tan\Big[{m}_2(\omega_0+&\delta\omega)y+{\Delta}_1^{(A)}(\omega_0+\delta\omega)\Big]\\
&\approx\tan\left[\left(\overline{m}_2+\delta m\right)y +\overline{\Delta}_1^{(A)}-\left(\delta m\right)x\right]\\ 
&=\tan\left[\overline{m}_2y+\overline{\Delta}_1^{(A)}+y(1-S)\delta m\right]\\
&=\frac{\tan\left[\overline{m}_2y+\overline{\Delta}_1^{(A)}\right]+\tan\left[y(1-S)\delta m\right]}
{1-\tan\left[\overline{m}_2y+\overline{\Delta}_1^{(A)}\right]\tan\left[y(1-S)\delta m\right]}\\
&=\tan\left[y(1-S)\delta m\right]\\
&\approx y(1-S)\delta m\ ;
\end{align*}
similarly, $\cot[\overline{m}_2y+\overline{\Delta}_1^{(B)} + y(1-S)\delta m]=-\tan[y(1-S)\delta m]\approx -y(1-S)\delta m$, since we can choose $|\delta m|y(1-S)\ll 1$ even if $|\delta m|y\gg1$ (small shell thickness).
Hence, by assuming a shell with high refractive index so that $|m_2|y\gg1$, the condition $S\approx1$ guarantees the validity of our analysis.
Notice, however, if one imposes $|\delta m|y\ll1$, this analysis is valid for every $0<S<1$. In particular, for $S\approx0$ [homogeneous sphere $(\varepsilon_2,\mu_2)$], we retrieve the results of Ref.~\cite{tribelsky}.
Substituting these approximations into Eqs.~(\ref{F1-te})--(\ref{G1-tm}) and redefining the auxiliary functions, we obtain:
\begin{align}
\mathcal{F}_1^{(A)}&\approx\frac{y^3}{3}\left[\frac{2{\widetilde{m}}_2}{y^2(1-S)}+\delta m\right]\ ,\label{F1-A-ap}\\
\mathcal{G}_1^{(A)}&\approx\frac{\widetilde{m}_2}{y^2(1-S)}-\delta m\ ,\label{G1-A-ap}\\
\mathcal{F}_1^{(B)}&\approx-\frac{y^3}{3}\left[\frac{\widetilde{m}_2}{1-S}-2\delta m\right]\ ,\label{F1-B-ap}\\
\mathcal{G}_1^{(B)}&\approx\frac{\widetilde{m}_2}{1-S}+\delta m \ .\label{G1-B-ap}
\end{align}

From Eqs.~(\ref{F1-A-ap}) and (\ref{G1-A-ap}), the constructive and destructive interferences in $|a_1|^2$ occur for $\delta m_+^{(A)}\equiv\widetilde{m}_2/[y^2(1-S)]$ and $\delta m_-^{(A)}\equiv-2\widetilde{m}_2/[y^2(1-S)]$, respectively.
Analogously, constructive and destructive interferences in $|b_1|^2$ occur for $\delta m_-^{(B)}\equiv-\widetilde{m}_2/(1-S)$ and $\delta m_+^{(B)}\equiv\widetilde{m}_2/[2(1-S)]$, respectively.
In particular, we have analytically demonstrated that both $|a_1|^2$ and $|b_1|^2$ have a Fano lineshape~\cite{tribelsky}: $|a_1|^2\propto(\rho + \beta)^2/(\rho^2+1)$, where $\rho=[(1+\alpha)\delta m - \alpha\delta m_+^{(A)}-\delta m_-^{(A)}]/[|\gamma|\sqrt{\alpha(3-\alpha)}]$, $\alpha=y^6/9$, $\beta=2\gamma/[|\gamma|\sqrt{\alpha(3-\alpha)}]$ and $\gamma = \delta m_-^{(A)}-\delta m_+^{(A)}$.
The $|b_1|^2$ profile is obtained from $|a_1|^2$ by replacing $\delta m_{\pm}^{(A)}$ with $\delta m_{\mp}^{(B)}$.
Assuming that only $\varepsilon_{2}$ depends on $\omega$ and $|\mu_{2{\rm r}}|\gg1$, we expect a Fano lineshape for both $|a_1|^2$ and $|b_1|^2$ as a function of $\varepsilon_{2}(\omega)$.

These asymmetric dipole resonances in total cross-sections are referred to as unconventional Fano resonances~\cite{tribelsky,tiago-pra}, which usually occur beyond the applicability of the Rayleigh approximation and are intrinsically different from the conventional Fano resonance that arises from the interference between different multipole orders in a specific scattering direction~\cite{luk,miroshnichenko}.
Here unconventional Fano resonances result from the interference of different EM modes excited inside the scatterer due to its high refractive index (hence, beyond the Rayleigh limit) with the same multipole moment $n=1$. 
Moreover, note that the asymmetry parameter $\beta$ has different signs in $|a_1|^2$ and $|b_1|^2$ profiles as a function of frequency [see the terms $\pm\delta m$ in Eq.~(\ref{F1-A-ap})--(\ref{G1-B-ap})].
This provides a configuration in which the destructive interferences of the electric and magnetic dipoles coincide without overlapping the scattering resonances.
Therefore, we have ``comblike'' electric and magnetic dipole resonances for $|m_1|x\ll1$, $|m_2|y\gg1$, and $y\ll1$, with the additional condition $S\approx1$ if $|\delta m|y\gg1$.
The analysis for the case $|\varepsilon_{2{\rm r}}|\gg1$, with $|\mu_{2{\rm r}}(\omega)|\ll1$, is completely analogous.

\subsection{Comblike resonances in core-shell particles}
\label{comblike}

We have demonstrated so far that it is possible to obtain, in the long-wavelength regime ($kb\ll1$), multiple Fano, comblike resonances in the scattering cross-section of concentric core-shell spheres. 
The basic conditions consist of a core in the Rayleigh limit $(|m_1|ka\ll1$, $ka\ll1$) and a concentric shell with both large refractive index $(|m_2|kb\gg1)$ and aspect ratio near to unity ($a\approx b$).
With a proper choice of parameters, one can obtain electric and magnetic Fano-like resonances with a scattering minimum at the same frequency (antiresonance or Fano-dip).
As we shall demonstrate, by making these approximations the Fano-dips of the electric and magnetic dipole resonances overlap without any additional assumption, giving rise to the unconventional Fano resonances. However, the approximation $S\approx 1$ (small shell thickness) is now mandatory. 

To identify the positions in frequency where these comblike resonances occur, we set $\widetilde{m}_2\to0$ for $|\mu_{2{\rm r}}|\to\infty$ (ferromagnetic limit), leading to $A_1\approx \tan(m_2x)$ and $B_1\approx-\cot(m_2x)$.
From Eqs.~(\ref{F1-te}) and (\ref{F1-tm}), the cloaking condition becomes $\tan[m_2y(1-S)]=0$, {\it i.e.}, $m_2y(1-S)=N\pi$, with $N=\pm1,\pm2,\ldots$
Note that we impose $S\approx1$ because $|m_2|y\gg1$.
In particular, this cloaking condition is readily obtained from Eq.~(\ref{F1-te}), but not from Eq.~(\ref{F1-tm}).
By showing that $\Delta_1^{(B)}\approx\tan^{-1}[\cot(m_2x)]=\pi/2-\cot^{-1}[\cot(m_2x)]=\pi/2-m_2x$ and using the trigonometric identity $\cot(\varphi_1+\varphi_2)=(\cot\varphi_1\cot\varphi_2-1)/(\cot\varphi_1+\cot\varphi_2)$, one can easily demonstrate that Eq.~(\ref{F1-tm}) also provides $\tan[m_2y(1-S)]=0$.
Explicitly, we have:
\begin{align*}
\cot\left[m_2y+\Delta_1^{(B)}\right]&\approx\cot\left[m_2y+\pi/2-m_2x\right]\\
&=\cot\left[m_2y(1-S)+\pi/2\right]\\
&=-\tan\left[m_2y(1-S)\right]\ .
\end{align*}
The condition $N=0$ ({\it i.e.}, $S=1$ since $|m_2|y\gg1$) leads to a homogeneous dielectric sphere $(\varepsilon_{1},\mu_{1})$, and, thereby, there is no cloaking.
These integer numbers $N$, which discretize EM cloaking for a certain set of parameters, can roughly be interpreted as indicating the ``cavity modes'' of the magnetic shell.
In fact, being $\lambda_2$ the wavelength inside a nanoshell with thickness $d\equiv b-a$, the EM cloaking occurs for $\lambda_2\approx2d/N$.
Note that this is the same condition for resonant modes in a Fabry-P\'erot cavity with $d$ being the separation distance between its mirrors. 
Since this analytical result is typical of resonant optical cavities, it suggests that the cloaking condition does not depend on the geometry of the particle (optical cavity), but rather the magnetic material parameters, shell thickness, and irradiation schemes. 
Analogously, similar results are obtained for the reciprocal configuration (high permittivity coatings) setting $1/\widetilde{m}_2\to0$, for $|\varepsilon_{2{\rm r}}|\to\infty$, where the same cloaking condition is achieved.  

Rewriting the cloaking condition, we explicitly have a mutual dependence between $\varepsilon_{2{\rm r}}$ and $\mu_{2{\rm r}}$, for $y\ll1$, $|m_1|x\ll1$, $|m_2|y\gg1$, and $S\approx1$:
\begin{align}
    {\varepsilon_{2{\rm r}}\mu_{2{\rm r}}}\approx\left[\frac{\pi N}{y(1-S)}\right]^2\ ,\quad N=\pm1,\pm2,\ldots\label{cloaking}
\end{align}
Note from Eq.~(\ref{cloaking}) that for $y\approx1$, large values of $\mu_{2{\rm r}}$ (or $\varepsilon_{2{\rm r}}$) are obtained for $S\approx 1$ (small shell thickness).
Besides, Eq.~(\ref{cloaking}) does not depend on $(\varepsilon_{1},\mu_{1})$, and this independence on the core material parameters is valid for $|m_1|x\ll1$.
This is of great interest, for instance, to cloak an arbitrary dielectric particle.
Also, due to the electric and magnetic Fano-like resonances, the cloaking frequencies given by Eq.~(\ref{cloaking}) are spectrally near to a pair of scattering resonance peaks. 
Each of these scattering, Fano-comb resonances are associated with a mode number $N$ and are referred to as morphology-dependent resonances of first order $(n=1)$. 

\section{Numerical results and discussions}
\label{numerical}

In the following we do not restrict ourselves to the small-particle limit, in which our previous analytical results are based, but we rather consider the exact expressions within the Lorenz-Mie theory, given by Eqs.~(\ref{Qsca})--(\ref{Bn}).
In particular, we consider a dielectric core with optical properties $(\varepsilon_{1{\rm r}}=10,\mu_{1{\rm r}}=1)$.
For the shell material, we consider a lossy Drude model for the relative permittivity, $\varepsilon_{\rm Drude}=\varepsilon_{\infty}-\omega_{\rm p}^2/[\omega(\omega+\imath\Gamma)]$, with the same parameters as used in Ref.~\cite{monticone2013} for aluminum-doped zinc oxide semiconductors: $\varepsilon_{\infty}=3.3$, $\omega_{\rm p}=2213.2$~THz, and $\Gamma=0.002\omega_{\rm p}$. 
The core radius is $a=100$~nm and the aspect ratio is $S=0.9$ ($b\approx 0.13\lambda$).
For the shell permeability, instead of assuming $\mu_{2{\rm r}}=1$ (nonmagnetic material), we investigate the cases $\mu_{2{\rm r}}=10^3$, $2\times10^3$, $5\times10^3$, and $10^4$.

\begin{figure}[htpb]
\centerline{\includegraphics[width=\columnwidth]{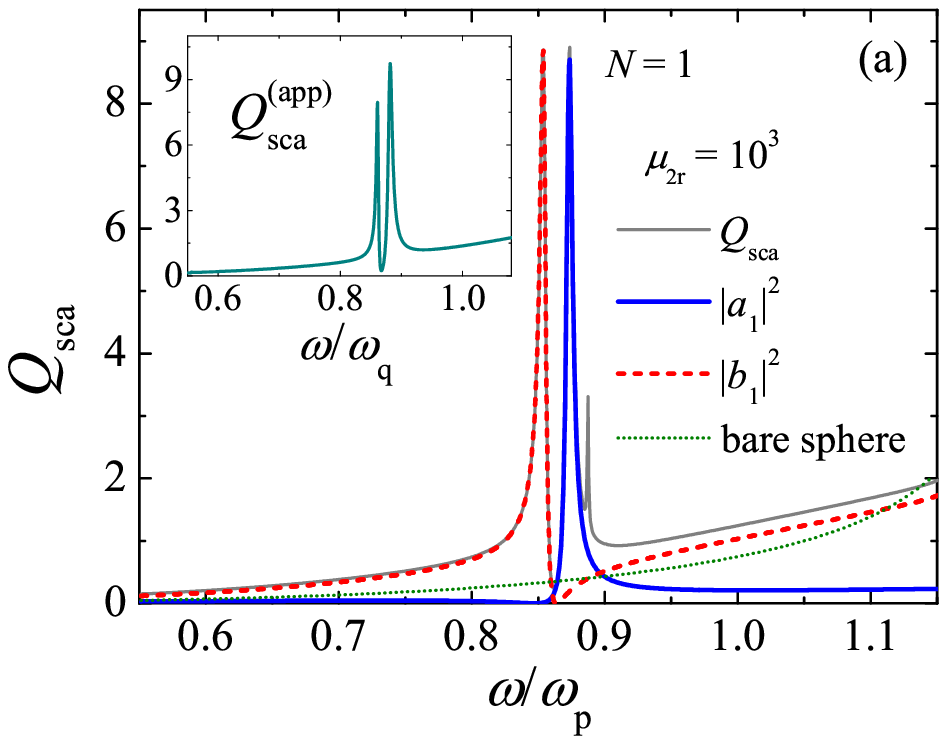}}
\centerline{\includegraphics[width=\columnwidth]{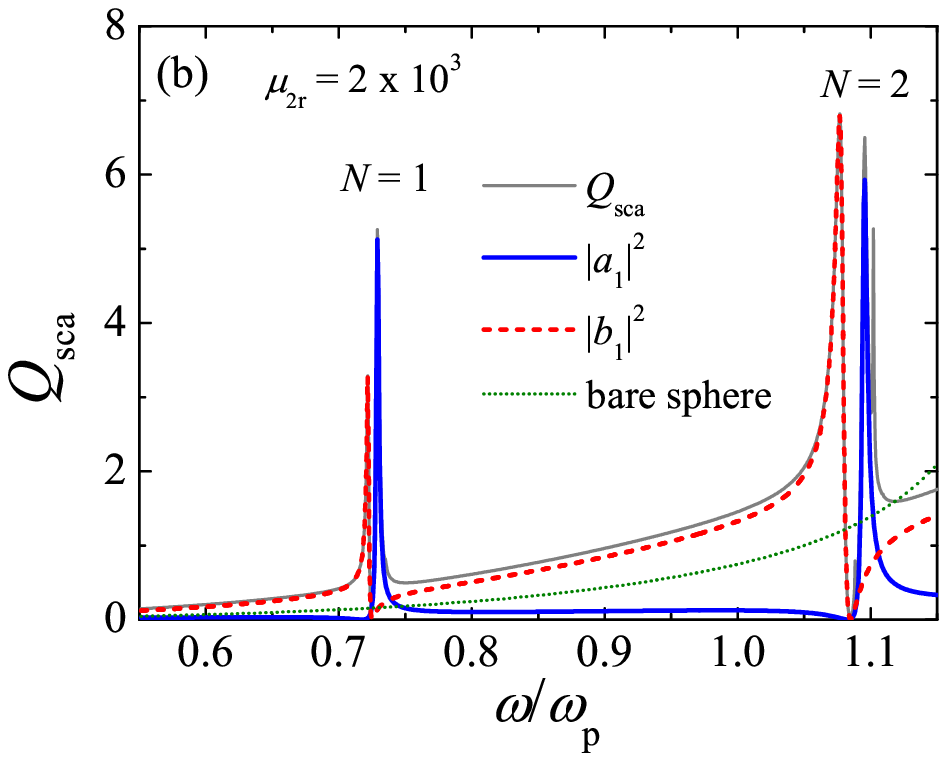}}
\caption{(Color online) Scattering efficiency $Q_{\rm sca}$ for a dielectric sphere $(\varepsilon_{1{\rm r}}=10,\mu_{1{\rm r}}=1)$ coated with a plasmonic, lossy magnetic shell $[\varepsilon_{2{\rm r}}(\omega)=\varepsilon_{\rm Drude},\mu_{2{\rm r}}\gg1]$.
The inner sphere has radius $a=100$~nm and aspect ratio $S=a/b=0.9$.
The dotted line represents the sphere without coating.
The cloaking occurs at $\varepsilon_{2{\rm r}}$ values given by Eq.~(\ref{cloaking}).
(a) $Q_{\rm sca}$ for $\mu_{2{\rm r}}= 10^{3}$. 
The $|a_1|^2$ and $|b_1|^2$ curves refer to the electric and magnetic dipole contributions within the exact Lorenz-Mie theory, respectively.
The inset is calculated from the dipole approximations (\ref{F1-te})--(\ref{G1-tm}), showing that $Q_{\rm sca}^{\rm (app)}$ is in good agreement with $Q_{\rm sca}$. 
(b) $Q_{\rm sca}$ for $\mu_{2{\rm r}}=2\times 10^{3}$.}\label{fig2}
\end{figure}

Figure~\ref{fig2} shows comblike, multiple Fano resonances in the scattering efficiency $Q_{\rm sca}$ for ${\rm Re}[\varepsilon_{2{\rm r}}(\omega)]>0$ and $\omega$ in the vicinity of $\omega_{\rm p}$, with $1.2<m_1x<1.7$ and $0.4<y<1$.
In Fig.~\ref{fig2}(a), we plot the scattering efficiency for $\mu_{2{\rm r}}=10^3$ ({\it i.e.}, $1.7<|m_2|y<48$) showing both the electric and magnetic dipole contributions with the corresponding antiresonance or Fano-dip in between.
The inset of Fig.~\ref{fig2}(a) shows the scattering efficiency $Q_{\rm sca}^{\rm(app)}$ calculated from the electric and magnetic dipole approximations, given by Eqs.~(\ref{F1-te})--(\ref{G1-tm}). 
As can it be confirmed by comparing the inset to the main plot of Fig.~\ref{fig2}(a), these approximations are in good agreement with the exact Lorenz-Mie theory, from which the main plot in Fig.~\ref{fig2}(a) is calculated.
However, from now on, all the calculations are performed using the exact expressions for the Lorenz-Mie coefficients~\cite{bohren}.
Increasing $\mu_{2{\rm r}}$ up to $2\times10^3$ ({\it i.e.}, $2.4<|m_2|y<67$), a second scattering dip appears in the same spectral frequency range, as evinced in Fig.~\ref{fig2}(b).
A very good estimate of the frequency position of these scattering minima is given by Eq.~(\ref{cloaking}), even though the Rayleigh and the small-particle limits are not entirely satisfied.
For this set of parameters, very efficient cloaking is achieved for a dielectric sphere $(\varepsilon_{1{\rm r}}=10,\mu_{1{\rm r}}=1)$ around $\omega\approx1.08\omega_{\rm p}$, for $N=2$ and $\mu_{2{\rm r}}=2\times10^3$.
We emphasize, nevertheless, that our aim is not to find an optimal set of parameters to cloak a dielectric particle~\cite{monticone-scatter-less}, but rather to achieve Fano-comb scattering resonances as a function of the magnetic permeability of the coating.

\begin{figure}[htpb]
\centerline{\includegraphics[width=\columnwidth]{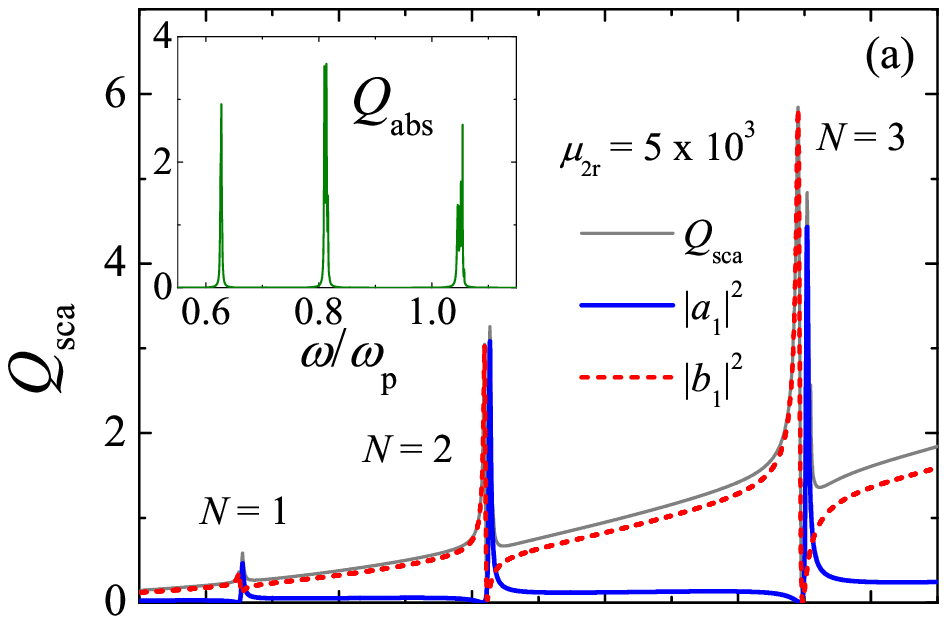}}
\vspace{-1.8cm}
\centerline{\includegraphics[width=\columnwidth]{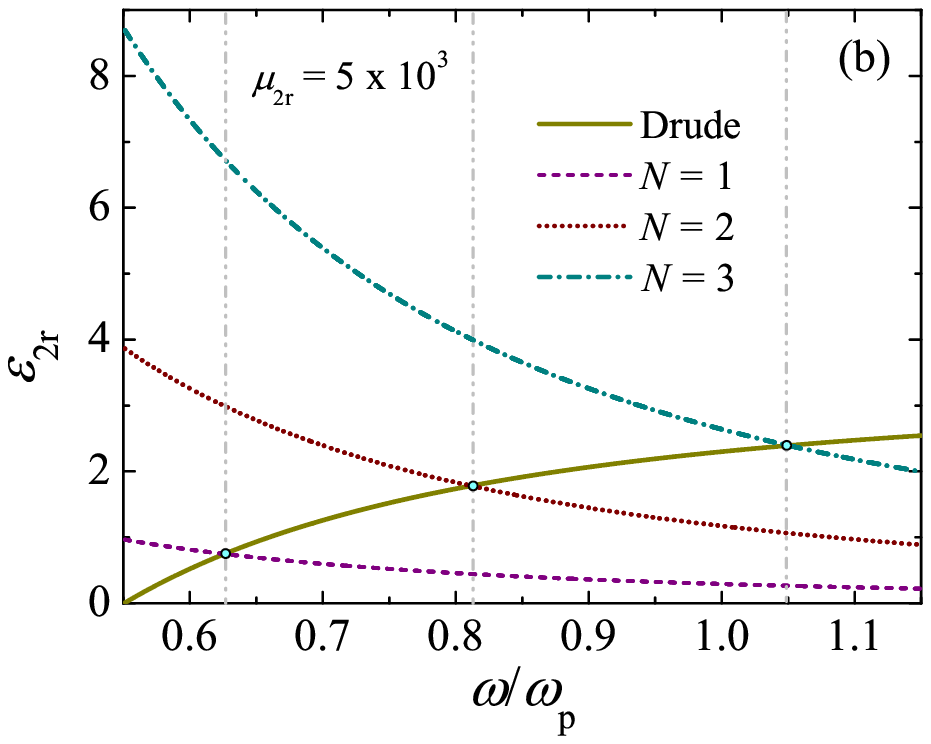}}
\caption{(Color online) Comblike scattering response for a dielectric sphere $(\varepsilon_{1{\rm r}}=10,\mu_{1{\rm r}}=1)$ coated with a plasmonic, lossy magnetic shell $[\varepsilon_{2{\rm r}}(\omega)=\varepsilon_{\rm Drude},\mu_{2{\rm r}}\gg1]$.
The inner sphere has radius $a=100$~nm and aspect ratio $S=a/b=0.9$.
The cloaking occurs at $\varepsilon_{2{\rm r}}$ values given by Eq.~(\ref{cloaking}).
(a) Scattering efficiency $Q_{\rm sca}$ and absorption efficiency $Q_{\rm abs}$ (inset) for $\mu_{2{\rm r}}=5\times 10^{3}$. 
(b) Shell permittivities $\varepsilon_{2{\rm r}}$ calculated from the Drude model and from Eq.~(\ref{cloaking}) for $\mu_{2{\rm r}}=5\times 10^{3}$ and $N=1$, $2$, and $3$.
The intersection between the vertical lines and the Drude curve corresponds to the localization, in frequency, of the Fano-comb.}\label{fig3}
\end{figure}

In Figs.~\ref{fig3}(a) and \ref{fig3}(b), we show the comblike scattering response for $\mu_{2{\rm r}}=5\times10^3$ and compare it to the analytical prediction given by Eq.~(\ref{cloaking}), which determines the values of $\varepsilon_{2{\rm r}}$ for which these multiple Fano resonances occur. As it can be seen from Fig.~\ref{fig3}, Eq.~(\ref{cloaking}) gives an excellent prediction for the frequency position of comblike resonances even for a lossy shell with $\varepsilon_{2{\rm r}}(\omega)>1$, $y\sim1$, and for higher orders of $N$. 
Figure~\ref{fig3}(b) corroborates that Eq.~(\ref{cloaking}) is valid even beyond the Rayleigh limit for the core and is robust against absorption, provided that $|m_2|y\gg1$ (with $|\mu_2|\gg|\varepsilon_2|$) and $S\approx1$.  
In particular, note that each value of $N$ replicates the same scattering profile of electric and magnetic dipole resonances, but with different amplitudes, and that they are approximately equally spaced in frequency.  
Also, the inset in Fig.~\ref{fig3}(a) shows high absorption in these antiresonance scattering points (Fano-dips).
This high absorption can be associated with resonances of the EM energy within the scatterer~\cite{tiago-joa,tiago-active}. 
In fact, according to Refs.~\cite{tiago-pra,miroshnichenko}, the Fano resonance in the scattering cross sections may lead to off-resonance field enhancement within the particle.

It is important to mention that the profile and the position of the comblike resonances can be externally tuned in engineered magnetic materials. 
Indeed, by applying an external magnetic field and/or varying the temperature, one can change the value of $\mu_{\rm 2}$.
Ferromagnetic materials, for instance, typically follow the Curie-Weiss law above their critical temperature $T_{\rm c}$~\cite{chen-mag}: $\mu_{2{\rm r}}\propto(T-T_{\rm c})^{-1}$.
Metamaterials made of ferroelectric compounds are examples of systems with tunable magnetic permeability at terahertz frequencies, with relatively low losses~\cite{metamu}. In these materials the tunability is achieved by varying the temperature, so that they could possibly be employed in an experimental verification of our findings.
This possibility of tuning the value of $\mu_{\rm 2}$ has influence not only on the resonance position but also on the number of resonances that compose the comblike scattering response at a given spectral range. 
This can be seen, {\it e.g.}, in Fig.~\ref{fig2}. 

We emphasize that, to the best of our knowledge, this tunability is not found in previous proposals involving comblike resonances in core-shell nanoparticles, which are based mainly on nonmagnetic materials~\cite{monticone2013,zayats}. 
In these previous systems, the position and profile of comblike resonances are determined {\it a priori} by the nanostructure design, either by varying the plasma frequencies with a gradient of doping level in multilayered spheres~\cite{monticone2013} or by constructing nonconcentric core-shell nanowires~\cite{zayats}. 
This fact, which distinguishes our proposal from the previous ones, allows for tunable, versatile comblike scattering response in concentric, single-layered core-shell nanoparticles.

\begin{figure}[htpb]
\centerline{\includegraphics[width=\columnwidth]{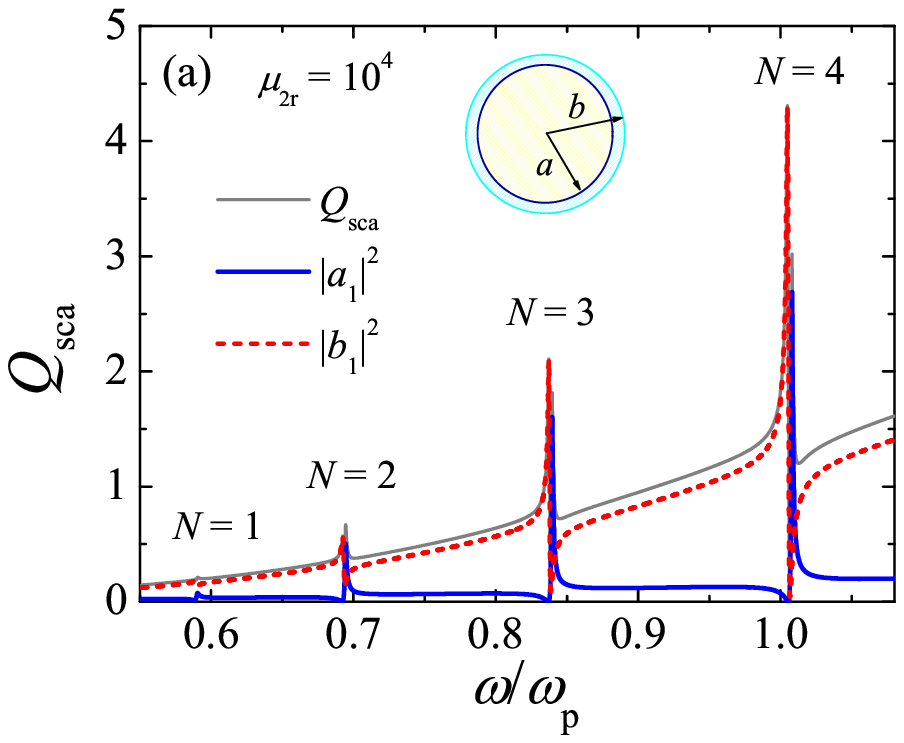}}
\centerline{\includegraphics[width=\columnwidth]{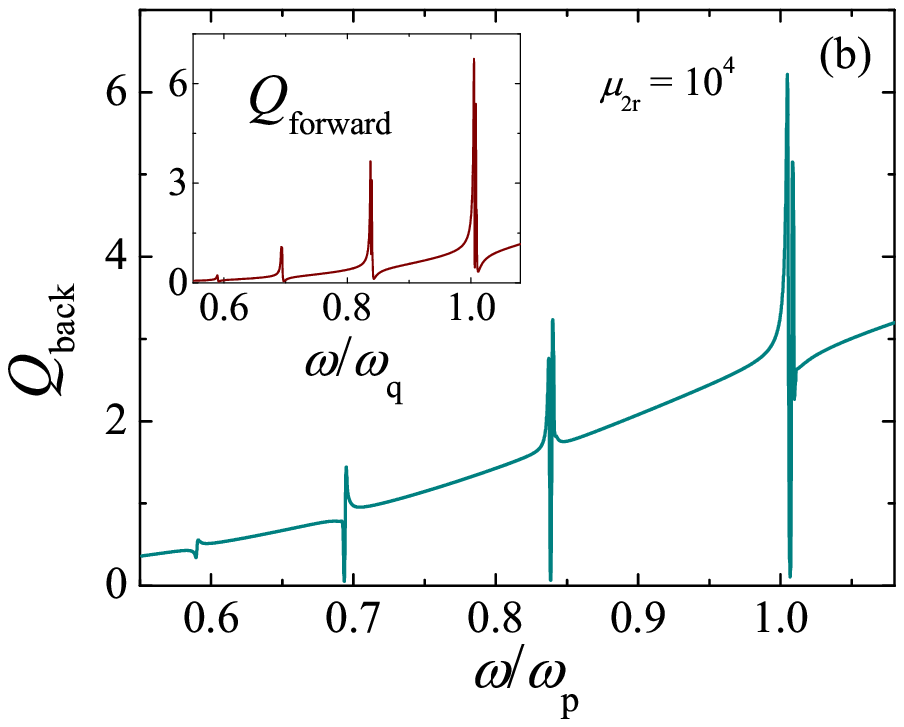}}
\caption{(Color online) Comblike scattering response for a dielectric sphere $(\varepsilon_{1{\rm r}}=10,\mu_{1{\rm r}}=1)$ coated with a plasmonic, lossy magnetic shell $[\varepsilon_{2{\rm r}}(\omega)=\varepsilon_{\rm Drude},\mu_{2{\rm r}}=10^4]$.
The inner sphere has radius $a=100$~nm and aspect ratio $S=a/b=0.9$.
The cloaking occurs at $\varepsilon_{2{\rm r}}$ values given by Eq.~(\ref{cloaking}).
(a) Scattering efficiency $Q_{\rm sca}$. 
(b) Differential scattering efficiencies $Q_{\rm back}$ and $Q_{\rm foward}$ (inset).}\label{fig4}
\end{figure}

In Fig.~\ref{fig4}(a), the shell permeability $\mu_{2{\rm r}}=10^4$ is one order of magnitude greater than in Fig.~\ref{fig2}(a), which implies that more scattering dips occur at the same spectral range. 
The differential efficiencies for the back and forward scattering~\cite{bohren}, respectively, $Q_{\rm back}=|\sum_{n=1}^{\infty}(2n+1)(-1)^n(a_n-b_n)|^2/y^2$ and  $Q_{\rm forward}=|\sum_{n=1}^{\infty}(2n+1)(a_n+b_n)|^2/y^2$, are plotted in Fig.~\ref{fig4}(b).
Due to the fact that the shell is magnetic ($\mu_{2{\rm r}} \neq 1$) light scattering is asymmetric with a backscattering dominance; indeed, $Q_{\rm back}$ has essentially the same profile as $Q_{\rm sca}$, as can be verified in Figs.~\ref{fig4}(a) and \ref{fig4}(b).
On one hand, the interference between electric and magnetic dipoles facilitates the suppression at the scattering dips of order $N$ in the backscattering direction.
On the other hand, the forward scattering efficiency $Q_{\rm forward}$, plotted in the inset of Fig.~\ref{fig4}(b), shows a strong suppression in the spectral frequency region between the $N$ and $N+1$ dips, and additional Fano-dips due to the conventional Fano resonance. 
This strong asymmetry between back and forward scattering may be exploited in applications. 
Indeed, the ultrasharp Fano-comb scattering suggests applications in optical tagging and signal processing, for instance.
In contrast to previous studies on multiple Fano resonances~\cite{monticone2013,zayats}, our approach not only allows one to analytically predict, by Eq.~(\ref{cloaking}), the frequency position where these resonances are expected to occur, but also allows for the possibility of tuning their positions via the external variation of the magnetic permeability of the nanoshell.

\section{Conclusions}
\label{conclusion}

Within the Lorenz-Mie theory, we have investigated the possibility of achieving multiple Fano resonances in the scattering response of dielectric spheres coated with a single-layer, concentric magnetic shell. 
We have derived, in the long-wavelength limit, explicit analytical conditions for the occurrence of comblike resonances in core-shell spheres. 
In particular, we have shown that nanoshells with high magnetic permeability or high electric permittivity values induce the formation of multiple Fano, comblike resonances composed of electric and magnetic dipole resonances. 
These Fano-comb resonances have been shown to follow a typical relation of a resonant optical cavity. 
As the condition for the occurrence of multiple Fano, morphology-dependent resonances explicitly depends on the magnetic permeability of the shell, we argue that both the position and profile of the comblike resonances could be tailored by an external magnetic field and/or by varying the temperature. 
Together with the simplicity of employing a single-layered core-shell nanoparticle, this tunability of the single scattering response distinguish our proposal from the previous ones considered so far, which are based on {\it e.g.} multilayered semiconductor materials  or nonconcentric core-shell particles. 
These singular scattering properties make the system proposed here a tunable, versatile optical device that may find applications in multifrequency biosensing, optical tagging and signal processing.

\section*{Acknowledgments}

The authors acknowledge the Brazilian agencies for support.
T.J.A. holds grants from Funda\c{c}\~ao de Amparo \`a Pesquisa do Estado de S\~ao Paulo (FAPESP) (2010/10052-0) and
A.S.M. holds grants from Conselho Nacional de Desenvolvimento Cient\'{\i}fico e Tecnol\'ogico (CNPq) (307948/2014-5).
F.A.P. thanks the Optoelectronics Research Centre and Centre for Photonic Metamaterials, University of Southampton, for the hospitality, and CAPES for funding his visit (BEX 1497/14-6). 
F.A.P. also acknowledges CNPq (303286/2013-0) for financial support.

\end{document}